\documentclass[prl,twocolumn,amsmath,amssymb,nofootinbib]{revtex4}

\usepackage{graphicx}
\usepackage{dcolumn}
\usepackage{bm}
\usepackage{epsfig}
\usepackage{graphicx,subfigure}

\usepackage{amsmath}
\usepackage{amsfonts}
\usepackage{amssymb}

\usepackage{amsmath}
\usepackage{amsfonts}
\usepackage{amssymb}
\usepackage{color}

\def\be{\begin{equation}}
\def\ee{\end{equation}}
\def\bea{\begin{eqnarray}}
\def\eea{\end{eqnarray}}

\def\ssdl{SS2$\ell$}

\def\slashchar#1{\setbox0=\hbox{$#1$}           
   \dimen0=\wd0                                 
   \setbox1=\hbox{/} \dimen1=\wd1               
   \ifdim\dimen0>\dimen1                        
      \rlap{\hbox to \dimen0{\hfil/\hfil}}      
      #1                                        
   \else                                        
      \rlap{\hbox to \dimen1{\hfil$#1$\hfil}}   
      /                                         
   \fi}

\begin{document}

\title{Same-Sign Dilepton Excesses and Light Top Squarks}

\author{Peisi Huang$^{a,b}$,  Ahmed Ismail$^{b,c}$, Ian Low$^{b,d}$ and Carlos E.~M.~Wagner$^{a,b,e}$ }

\affiliation{
\mbox{{$^a$Enrico Fermi Institute, University of Chicago, Chicago, IL 60637, USA}}\\
\mbox{$^b$ High Energy Physics Division, Argonne National Laboratory, Argonne, IL 60439, USA}\\
\mbox{$^c$Department of Physics, University of Illinois, Chicago, IL 60607, USA }\\
\mbox{$^d$ Department of Physics and Astronomy, Northwestern University, Evanston, IL 60208, USA} \\
 \mbox{$^e$Kavli Institute for Cosmological Physics, University of Chicago, Chicago, IL 60637, USA}}

\begin{abstract}
Run 1 data of the Large Hadron Collider (LHC) contain excessive events in the same-sign dilepton channel with $b$-jets and missing transverse energy (MET), which were observed by five separate analyses from  ATLAS and CMS collaborations. We show that these events could be explained by direct production of top squarks (stops) in supersymmetry. In particular, a right-handed stop with a mass of 550 GeV decaying  into 2 $t$ quarks, 2 $W$ bosons, and MET could fit the observed excess without being constrained by other direct search limits from Run 1. We propose  kinematic cuts at 13 TeV to enhance the stop signal, and estimate that stops could be discovered with 40 fb$^{-1}$ of integrated luminosity at Run 2 of the LHC, when considering only the statistical uncertainty.

\end{abstract}

\maketitle

\noindent {\bf Introduction} -- Run 1 of the LHC culminated in the discovery of the Higgs boson at 125 GeV. The initial indication of an excess in the diphoton channel by both the ATLAS and the CMS collaborations in December of 2011 \cite{ATLAS:2012ad,Chatrchyan:2012twa} set the stage for the celebrated announcement  in July of 2012 \cite{Aad:2012tfa,Chatrchyan:2012ufa}. Post Higgs discovery, the most important question is, naturally, where is  the new physics beyond the Standard Model (SM)? 

In this work we report on an excess of events in the same-sign dilepton (\ssdl) channel with $b$-jets and MET that were observed by five analyses from ATLAS and CMS using Run 1 data. These analyses are summarized below:
\begin{itemize}

\item CMS \ssdl\ SUSY Search \cite{Chatrchyan:2013fea}: This is a cut-and-count analysis. In the signal region ``SR24,'' defined as \ssdl, $N_{\rm b-jets} \ge 2$, $N_{\rm jets}\ge 4$, $E_{\rm T}^{\rm miss}\in [50,120]$ GeV, and $H_{\rm T}\ge 400$ GeV, the expected number of events is $4.4\pm 1.7$, in the region of $p^\ell_{\rm T}\ge 10$ GeV,  and $2.8\pm 1.2$ in the region of of $p^\ell_{\rm T}\ge 20$ GeV. The observed number of events are 11 and 7, respectively. No $p$-value is given. 

\item ATLAS \ssdl\ SUSY Search \cite{Aad:2014pda}: This is a cut-and-count analysis. In the signal region ``SR1b,'' defined as \ssdl, $N_{\rm b-jets} \in [1,2]$, $N_{\rm jets}\ge 3$, $E_{\rm T}^{\rm miss}\ge 150$ GeV,  $m_{\rm T}\ge 100$ GeV and $m_{\rm eff} \ge 700$ GeV, the number of expected background events is $4.7\pm 2.1$ while the observed number of events is 10. The $p$-value is 0.07.

\item CMS \ssdl\ ttH Search \cite{Khachatryan:2014qaa}: This is a multivariate analysis based on the Boosted Decision Tree (BDT). The \ssdl\ signal region is defined as \ssdl\ with $p_{\rm T}\ge 20$ GeV, $N_{\rm jets}\ge 4$ and $N_{\rm b-jets}\ge 1$ with $p_{\rm T}\ge 25$ GeV. The best fit signal strength $\mu$, in units of the SM ttH signal strength, is $5.3^{+2.1}_{-1.8}$.

\item ATLAS \ssdl\ Exotica Search \cite{Aad:2015gdg}: This is a cut-and-count analysis. The SRVLQ6/SR4t3 signal region requires \ssdl\ with possible additional leptons, $N_{\rm jets}\ge 2$, $N_{\rm b-jets}=2$, $H_{\rm T}\ge 700$ GeV and $E_{\rm T}^{\rm miss}\ge 100$ GeV. The expected number of events is $4.3\pm 1.1\pm 1.1$ and the observed number is 12. The $p$-value is 0.029.

\item ATLAS \ssdl\ ttH Search \cite{atlastth}: This is a cut-and-count analysis. The $2\ell 0 \tau_{\rm had}$ category requires \ssdl\ with $p_{\rm T}\ge 25 (20)$ GeV for the (sub)leading lepton and at least 4 reconstructed jets, at least one of which must be $b$-tagged. The observed signal strength is $\mu=2.8_{-1.9}^{+2.1}$.

\end{itemize}

What emerges from these observations is that there is a mild excess in the \ssdl\ channel with $b$-jets and MET in the LHC Run 1 data. While it is difficult to estimate the overall significance of the excess, and the SUSY search excesses are in different MET regions, it is worth noting that the CMS ttH search and the ATLAS Exotica search both reported a significance of 2$\sigma$ level or higher.

In what follows we will assume the \ssdl\ excess is due to physics beyond the SM and investigate scenarios which could potentially explain the excess. For simplicity we choose to base our simulations on the CMS \ssdl\ ttH search in Ref.~\cite{Khachatryan:2014qaa}, which provides a best fit signal strength. Specifically, we will normalize the signal strength in our new physics benchmarks to the SM ttH signal strength in this analysis. We anticipate that some, although not all, of the systematic uncertainties would cancel in this procedure. \\

\noindent {\bf General Classification} -- The excess in the \ssdl\ channel can be broadly characterized as
\be
2t + 2W + X
\ee
where $X$ contains additional particles. If $X$ = MET, then the final state contains 2 $b$-jets exclusively. One canonical example is pair production of new heavy particles decaying into a top quark, a $W$ boson, and MET. 

It is worth stressing that the assumption of $X$ = MET could be relaxed. For example, $X$ could contain, in addition to MET, accompanying visible particles such as $b$-jets, giving rise to final states with 3 or more $b$-jets. One possibility would be the production of four top quarks.

In this work we will adopt the simplifying hypothesis that $X$ = MET and focus on new physics contributing to the final states
\be
2t + 2W + {\rm MET} \ ,
\ee
leaving the more complicated scenarios for future work. While MET is normally attributed to the existence of a stable neutral particle, there could be accompanying soft, and possibly charged, particles that also escape detection. This is the scenario that we will employ in the case of top squark decays in supersymmetry.

One possibility to explain the \ssdl\ excess, without invoking the existence of new particles, is that the excess could be due to a modified Higgs coupling to the SM top quark, resulting in an enhanced tt(H$\to$ multileptons) production. There are two potential problems with this scenario: 1) Run 1 analyses do not exhibit similar enhancement in the tt(H$\to b\bar{b}$) channel \cite{Khachatryan:2014qaa, Aad:2015gra}, although the present uncertainty is quite large and an enhancement in the $b\bar{b}$ channel cannot be excluded with confidence yet, and 2) the gluon fusion production of the Higgs would  need to be enhanced at a similar level as the ttH enhancement, since in the SM the gluon fusion process is directly proportional to the top Yukawa coupling. Again this does not seem to be supported by global fits of Higgs data in Run 1 \cite{atlascombined,Khachatryan:2014jba}.

Therefore, we will pursue the possibility that the \ssdl\ excess is due to pair production of new colored particles, which  proceeds through identical decay chain. Postulating the existence of a stable neutral particle ${N}$, of arbitrary spin, the electric charge of the new particle could be classified. In all cases, the new colored particles could be a scalar, a fermion, or a vector boson, depending on the spin of $N$. The possibilities are

\begin{itemize}

\item A charge-($-1/3$) new particle ${\cal B}\to t + W^- + {N}$. A scalar example would be the bottom squark (sbottom) $\tilde{b}$ in supersymmetry decaying into $t+(\tilde{\chi}_1^-\to W^-\tilde{\chi}_1^0)$ \cite{Chatrchyan:2013fea,Aad:2014pda}. ${\cal B}$ could also be a vector-like fermion decaying into $t + (W^-_H\to W^- + A_H)$ as in  littlest Higgs theories with T-parity \cite{Cheng:2003ju}, where $W_H$ is a heavy cousin of the  $W$ boson and $A_H$ is the lightest T-odd particle.

\item A charge-(+2/3) new particle ${\cal T}\to t + W^\pm + C^\mp$, where $C^\pm$ is a heavy charged particle that is nearly degenerate with $N$ and subsequently decays into $N$ + soft charged particles. In this case $C^\pm$ will manifest itself as MET in the detector.  This case will be discussed in detail in the next Section.

\item A charge-(+5/3) new particle ${\cal X}_{5/3}\to t + W^+ + N$. One closely related example in the literature is the charge-(+5/3) $X_{5/3}$ fermion in composite Higgs models, which decays into $t+W^+$ \cite{Contino:2008hi}. However, in this case the MET arises solely from  the neutrino in the $W$ decay.

\end{itemize}

For all possible spin quantum numbers of the new particles involved, one could construct ``simplified models'' where the decay branching ratio (BR) into the desired final states is 100\%. In a complete model, however, this is sometimes difficult to achieve. For example, the sbottom in supersymmetry has two possible decay channels:
\bea
\tilde{b}&\to& t+\tilde{\chi}_1^- \to t+(W^-+\tilde{\chi}_1^0) \ ,\nonumber \\
\tilde{b}&\to& b+\tilde{\chi}_1^0 \ . \nonumber 
\eea
Only the former gives the \ssdl\ signature, which comes from the left-handed component of $\tilde{b}$. While the desired channel can be made to dominate in the case of Higgsino-like $\tilde{\chi}_1^\pm$ and bino-like $\tilde{\chi}_1^0$, the decays of the $\tilde{t}_L$ must then also be considered. The left-handed stop would preferentially decay to neutral Higgsinos, which would then decay to the $\tilde{\chi}_1^0$. The spectrum would give additional top pair production, and for sufficiently small mass splittings the Higgsino decays would produce off-shell Z bosons, leading to an edge in the dilepton mass distribution that would be smaller than that observed by CMS \cite{Huang:2014oza}. Such a case is beyond the scope of this work, but would be interesting to study further. In what follows we will consider a realistic model of right-handed stop decays in supersymmetry, where the branching fraction into the \ssdl\ final state can be very significant without additional complications.\\

\noindent {\bf A Realistic Model: The Stop} -- In supersymmetry, stops are particularly important because of their roles in raising the tree-level mass of the lightest CP-even Higgs as well as stabilizing the Higgs mass. (See, for example, Ref.~\cite{Carena:2002es} and references therein.) Here, we outline a viable scenario in the MSSM through which stop pair production can produce extra \ssdl\ events without being constrained by existing experimental searches. Given the signature outlined for a charge-(+2/3) particle ${\cal T}$ in the previous section, we will consider the following decay chain
\be
\label{eq:stopR}
\tilde{t}_R \to t + \tilde{B} \to t + (\tilde{W}^\pm + W^\mp)
\ee
where $\tilde{t}_R$ is the right-handed stop, $\tilde{B}$ is the bino and $\tilde{W}^\pm$ is the charged wino. In particular we assume that the lightest supersymmetric particle (LSP) is the neutral wino, which is nearly degenerate with the $\tilde{W}^\pm$ in mass. The charged wino will then decay into the LSP and soft charged particles, resulting in MET in the collider detector. In terms of mass eigenstates, the decay BR of the lightest stop ($\tilde{t}_1$) into top + second neutralino ($\tilde{\chi}^0_2$) can be  quite large as long as  $\tilde{t}_1$ is mostly right-handed and $\tilde{\chi}^0_2$ is bino-like.  Decays of $\tilde{\chi}^0_2$ into the wino-like $\tilde{\chi}^\pm_1$ and $W^\mp$  can also be dominant if $m_{\tilde{\chi}^0_2} - m_{\tilde{\chi}^0_1} < m_H=125$ GeV, so as to suppress the decay $\tilde{\chi}^0_2\to \tilde{\chi}^0_1+H$.

Given these considerations, we choose the following spectrum in the MSSM:

\begin{itemize}

\item A right-handed $\tilde{t}_1$ with mass $550$ GeV.

\item A bino-like $\tilde{\chi}^0_2$ with mass $340$ GeV.

\item Wino-like $\tilde{\chi}_1^\pm$ and $\tilde{\chi}^0_1$ with nearly degenerate mass $260$ GeV.

\end{itemize}
All other superpartners can be heavier than 1 TeV and decouple from the effective theory at the weak scale.  The stop then decays as shown in Eq.~(\ref{eq:stopR}). We note that there are no dedicated searches for stops in this particular channel giving rise to the \ssdl\ excess. However, the same final states have been looked for in the context of sbottom searches. One example is SUSY searches in three leptons and at least one $b$-jet in Ref.~\cite{cms3l}. The limit, however, disappears when $m_{\tilde{\chi}^0_1} \agt 240$ GeV for $m_{\tilde{\chi}^\pm_1}/m_{\tilde{\chi}^0_2}\le 0.8$, thus motivating our choice of 260 GeV mass for the LSP. The bino mass is chosen so that the wino-bino mass difference is smaller than the Higgs mass in order to suppress the bino decays into $\tilde{\chi}_1^0+H$. With these choices, other searches for 0 or 1 lepton, ($b$-)jets and MET are not expected to constrain our spectrum. We note that heavy left-handed stops, with soft SUSY-breaking masses of several TeV and similarly sized A-terms, can provide sufficient corrections to reproduce the 125 GeV Higgs mass without affecting the low energy spectrum we consider.

Disappearing track searches can in principle probe the wino-like $\tilde{\chi}^\pm_1$, but current bounds can be easily evaded. For our 260 GeV mass choice in the pure wino limit, the mass splitting between the $\tilde{\chi}^\pm_1$ and $\tilde{\chi}^0_1$ is roughly 160 MeV \cite{Ibe:2012sx}, which is near the current CMS limit \cite{CMS:2014gxa}. However, a small amount of Higgsino mixing can signficantly increase the mass splitting. For the physical masses above with $\mu$ = 1 TeV, SOFTSUSY \cite{Allanach:2001kg} predicts a $\tilde{\chi}^\pm_1-\tilde{\chi}^0_1$ splitting of 240 MeV, more than enough to avoid the disappearing track bound, and SDECAY \cite{Muhlleitner:2003vg} gives BR($\tilde{t}_1 \to t + \tilde{\chi}^0_2$) = 93\%. For simplicity, in what follows we assume that all branching ratios are 100\%.

We simulate stop pair production events for the spectrum above, as well as SM ttH, using \verb+MadGraph5_aMC@NLO+ \cite{Alwall:2014hca}, Pythia 6.4 \cite{Sjostrand:2006za} and Delphes 3 \cite{deFavereau:2013fsa} with anti-$k_T$ jet clustering using FastJet \cite{Cacciari:2011ma, Cacciari:2008gp}. Throughout, we normalize to cross sections from Ref.~\cite{Heinemeyer:2013tqa} and Ref.~\cite{Kramer:2012bx, Borschensky:2014cia} for SM ttH and for direct stop productions, respectively. We perform a cut-and-count simplification of the CMS ttH analysis, by implementing the signal selection cuts of the \ssdl\ search region in Ref.~\cite{Khachatryan:2014qaa} without modeling the BDT analysis. We find that the stop signal yield, in units of the SM ttH expectation, is $\mu_{\tilde{t}}({\rm 8\ TeV}) = 1.83$, giving rise to a total signal strength
\be
 \mu({\rm 8 \ TeV})=2.83\ ,
 \ee
after adding in the SM ttH contribution. This value is about 1.5$\sigma$ below the CMS central value \cite{Khachatryan:2014qaa} and in nearly perfect agreement with the ATLAS central value \cite{atlastth}.

Since the CMS and ATLAS ttH analyses also provided a best-fit signal strength in the trilepton ($3\ell$) category, as a check for the stop scenario we implemented the selection cuts of the CMS ttH $3\ell$ analysis in Ref.~\cite{Khachatryan:2014qaa} and found the signal strength to be
\be
\mu_{3\ell}({\rm 8 \ TeV)} = 2.1 \ ,
\ee
again in good agreement with the ATLAS and CMS  $3\ell$ signal fits, which are $2.8^{+2.2}_{-1.8}$ \cite{atlastth} and $3.1^{+2.4}_{-2.0}$ \cite{Khachatryan:2014qaa}, respectively. There is also the $4\ell$ category which has a rather large uncertainty and is not considered here.\\

\noindent {\bf Run 2 Projections} -- We now turn to the prospect of observing the \ssdl\ excess at the LHC Run 2. The first important observation is that the  cross sections for the ttH and the direct stop productions increase at different rates in going from 8 TeV to 13 TeV, as can be seen in Table \ref{table:xsection}. In addition, the dominant background in the SM comes from ttV productions, where V=$W/Z/\gamma^*$. All three processes are produced through the gluon initial states, therefore we expect the heaviest final state to gain the most in going from 8 TeV to 13 TeV. In other words, the increase in the rate for stop production would be larger than the ttH production, which in turn would outgrow the dominant SM background. As a result, if the \ssdl\ excessive events are due to stop production, the enhancement relative to the SM ttH signal strength in the \ssdl\ category should grow in going from 8 TeV to 13 TeV. Indeed, using the same selection cuts as in 8 TeV, we find the stop benchmark gives $\mu_{\tilde{t}}({\rm 13\ TeV}) = 2.69$ and hence the total signal yield
\be
\label{eq:estmu}
\mu({\rm 13\ TeV}) =3.69 \ ,
\ee
relative to the SM ttH signal strength.

\begin{table}[t]
\centering
\begin{tabular}{c c c c}
\hline\hline
 &$ \sigma$(8 TeV)  & $\sigma$(13 TeV) &  Ratio(13 TeV/8 TeV)   \\
 \hline\hline
$\sigma(pp\to {\rm ttH})$&   { 129 fb} & { 509 fb} & 3.9 \\
$\sigma(pp\to \tilde{t}_1\tilde{t}_1^*) $ & { 45 fb} & { 296 fb}& 6.6 \\
 \hline
\end{tabular}
 \caption{SM ttH  \cite{Heinemeyer:2013tqa} and direct stop production cross sections \cite{Kramer:2012bx, Borschensky:2014cia}.}
\label{table:xsection}
\end{table}

\begin{figure*}[!t]
  \begin{center}
    \includegraphics[width=0.42\textwidth]{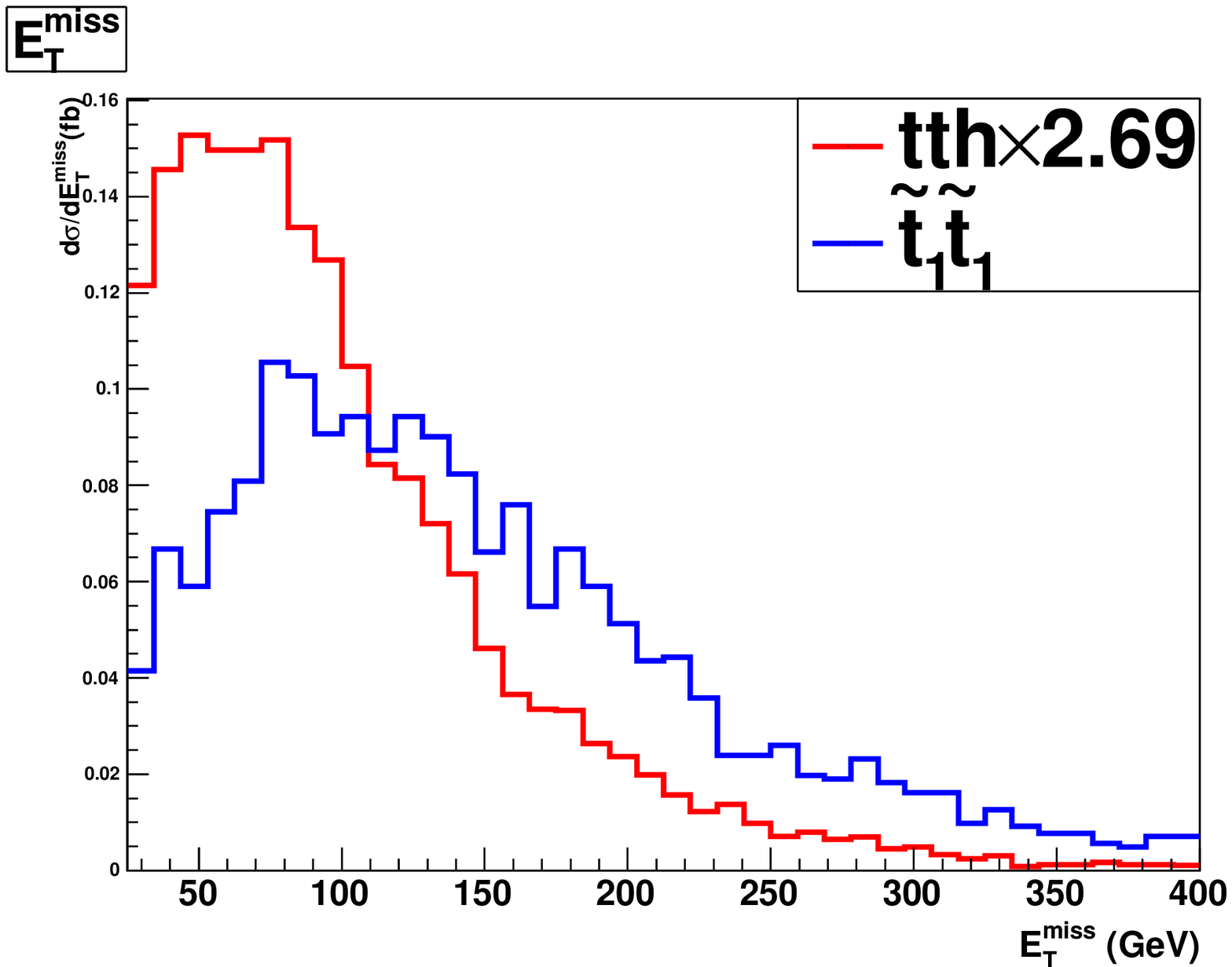}
        \includegraphics[width=0.42\textwidth]{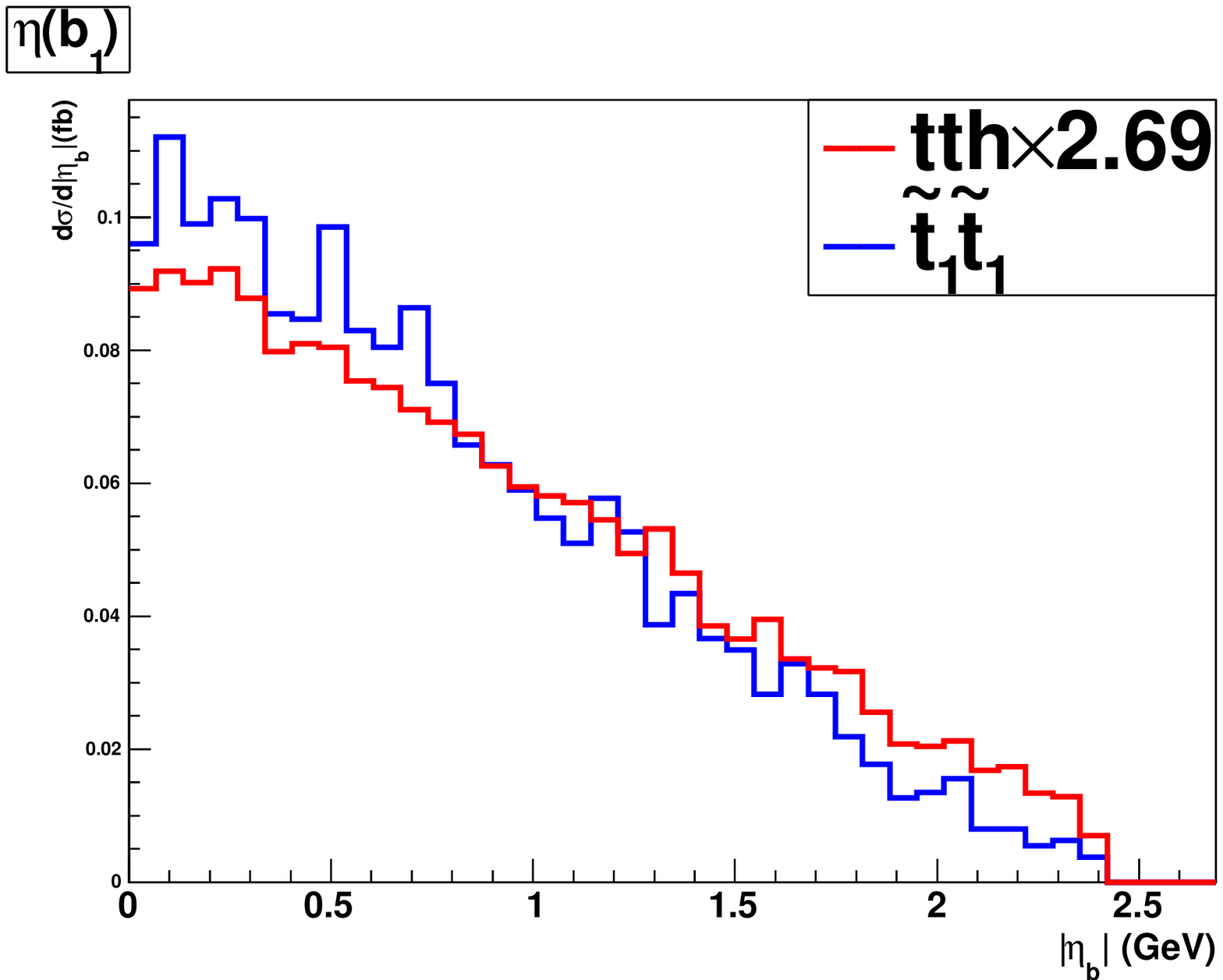}
    \caption{\label{fig:13tev}Kinematic distributions of $\eta_b$ and $E_{\rm T}^{\rm miss}$ of ttH and stop events at the 13 TeV LHC. The distributions in each plot have the same normalization.}
    \end{center}
\end{figure*}

In addition to this enhanced signal yield, some kinematic distributions of the decay products are different between ttH and stop production. In Fig.~\ref{fig:13tev} we show distributions of the MET ($E_{\rm T}^{\rm miss}$) and the pseudo-rapidity of the $b$-jets ($\eta_b$) at the 13 TeV LHC. (Distributions at the 8 TeV LHC look similar.) As expected, the $H_{\rm T}$ distribution from stop pair production extends further out than that for ttH. Also, even though our stop spectrum is somewhat compressed, the stop events tend to have more missing energy than those from ttH. Finally, we note that in the stop events, the $b$-jets are more centrally produced. This is a consequence of the two tops in the final state coming from the decay of stops, which tend to be produced with little momentum, rather than from the production of ttH, which tends to be more forward due to the $t$-channel kinematics.

These observations motivate the following cuts to discriminate stops from ttH events, which we impose in addition to the existing cuts of the CMS ttH \ssdl\ search:

\begin{itemize}

\item MET $> 125$ GeV

\item $|\eta_b| < 1$ for one (medium) or two (loose) b-jets

\end{itemize}
At 13 TeV with these additional cuts, the expected yield from our stop benchmark grows from 3.69 in Eq.~(\ref{eq:estmu}) to
\be
\label{eq:newcuts}
{\mu}({\rm 13\ TeV)} = 6.94
\ee
in units of the SM ttH strength. On the contrary, if the excess were due to an enhanced top Yukawa coupling, signal strength would not change in going from 8 to 13 TeV, modulo experimental uncertainties.

In our simulations, the number of SM ttH events passing our additional cuts at 13 TeV is about $76\%$ of SM ttH passing the original CMS cuts at 8 TeV in Ref.~\cite{Khachatryan:2014qaa}. Given this consideration, and making the conservative assumption that the SM ttV background grows at the same rate as ttH in going from 8 to 13 TeV, we estimate the stop signal strength in Eq.~(\ref{eq:newcuts}) would be discoverable above SM ttH with 40 fb$^{-1}$ of Run 2 data. In this estimate we consider only the statistical uncertainty and have not included systematic errors \cite{Curtin:2013zua, Khachatryan:2014qaa}, but expect that their relative influence may be reduced by tightening the illustrative cuts considered here. This result motivates a more complete investigation  by experimental collaborations.

In addition to the \ssdl\ channel, we note that the bino may decay to either sign of charged wino, and so it is possible to get the stop decay products $t\bar{t} + W^\pm W^\pm + {\rm MET}$. In principle, this can lead to final states with three or more same-sign leptons, where the SM background would be extremely low. With the 40 fb$^{-1}$ of Run 2 data that would be needed to conclusively discover our stop spectrum in \ssdl\ + $b$-jets + MET, we expect approximately 5 $\ell^\pm\ell^\pm\ell^\pm$ events. As for the \ssdl\ channel of the CMS ttH search, the largest background to this same-sign trilepton signal would likely be non-prompt leptons, and a simple estimate using typical fake rates gives ${\cal O}(0.1)$ events for the same luminosity. Should the \ssdl\ excess persist without a corresponding signal in same-sign trileptons, other topologies that we have discussed, such as sbottoms, could prove useful in providing an explanation. Other conventional search channels are less likely to be competitive with the \ssdl\ signature we have considered. For example, one lepton searches would have much higher backgrounds. Also, (non-same-sign) trilepton searches suffer sufficiently from the low BR that at 8 TeV the sensitivity \cite{cms3l} is less than that for \ssdl\ searches, and we expect this trend to continue at 13 TeV.\\

\noindent {\bf Outlook} -- On the verge of LHC Run 2, clear signs of physics beyond the SM have thus far remained elusive. If new phenomena have been present in Run 1 data, their signatures have been at or beyond the reach of existing searches, and potential hints of novel physics should be scrutinized more carefully. Here we have identified such a possibility in events with \ssdl, $b$-jets and MET. We have outlined potential explanations for this excess, and focused on a supersymmetric scenario where stop decays could provide a source for \ssdl\ events.

For our stop scenario, we have considered constraints from both supersymmetric and Higgs searches, presenting a spectrum which remains viable with current data. Early 13 TeV data would show clear signatures of this spectrum. We have described kinematic cuts that could help enhance the stop pair production over the ttH events. We have also highlighted a completely new search channel with very low background that would provide a separate probe of our model. We look forward to elucidating these prospects at Run 2.

From the model building perspective, it would be interesting to construct UV completions giving rise to the particular stop spectrum that we considered in the benchmark. In this regard we note that the right-handed stop is typically the lightest squark at the weak scale when starting the renormalization group evolution from a universal value at the high scale. On the other hand, the LSP in our benchmark is a pure wino with a mass that is too light to achieve the correct dark matter relic density; additional contributions to the relic density, e.g. from axions, or non-thermal dark matter production will be necessary. We expect to return to these aspects of a possible UV completion in a future work.

\begin{acknowledgments}
\noindent
{\bf Acknowledgements}: We thank Stefania Gori for collaborations in the early stage of this work. Useful discussions with Frank Golf, Ben Hooberman, Aurelio Juste, Kevin Lannon, Jeremy Love, Sasha Paramonov and Michael Ramsey-Musolf are gratefully acknowledged. Work at ANL is supported by the U.S. Department of Energy under grant No. DE-AC02-06CH11357. A.I. is partially supported in part by the U.S. Department of Energy under grant DE-FG02-12ER41811. P.H. is partially supported by U.S. Department of Energy Grant DE-FG02-04ER41286. I.L. is partially supported in part by the U.S. Department of Energy under grant No. DE-SC0010143.
\end{acknowledgments}


\end{document}